\documentclass[prb]{revtex4}

\usepackage{graphicx}
\usepackage{dcolumn}
\usepackage{bm}

\begin{document}

\title{\bf Experimental study of weak antilocalization effect
in a high mobility InGaAs/InP quantum well}
\author{S.A. Studenikin \footnote{sergei.studenikin@nrc.ca},
P.T. Coleridge, N. Ahmed$^\# $, P. Poole and A. Sachrajda}
\affiliation{Institute for Microstructural Sciences, National
Research Council of Canada, Ottawa, Ontario, K1A OR6, Canada \\
$^\# $ Department of Physics, University Brunei Darussalam,
Brunei Darussalam}
\date{25 February 2003}
\begin{abstract}

The magnetoresistance associated with quantum interference
corrections in a high mobility, gated InGaAs/InP quantum well
structure is studied as a function of temperature, gate
voltage, and angle of the tilted magnetic field. Particular
attention is paid to the experimental extraction of
phase-breaking and spin-orbit scattering times when weak anti-
localization effects are prominent. Compared with
metals and low mobility semiconductors the characteristic
magnetic field $B_{tr} = \hbar/4eD \tau$ in high mobility
samples is very small and the experimental dependencies of the 
interference effects extend to fields several hundreds of
times larger.  Fitting experimental results under
these conditions therefore requires theories valid for
arbitrary magnetic field. It was found, however, that such a
theory was unable to fit the experimental data without
introducing an extra, empirical, scale factor of about 2. 
Measurements in tilted magnetic fields and as a function of
temperature established that both the weak localization and the
weak anti-localization effects have the same, orbital origin.
Fits to the data confirmed that the width of the low field
feature, whether a weak localization or a weak
anti-localization peak, is determined by the phase-breaking
time and also established that the universal (negative)
magnetoresistance observed in the high field limit is 
associated with a temperature independent spin-orbit scattering
time.

\end{abstract}

\pacs{71.30.+h,  72.20.-i, 73.20.Dx}

\keywords{weak localization, antilocalization, phase-breaking
time, spin-orbit coupling, magnetoresistance, quantum wells,
magnetotransport}

\maketitle
\newpage

\subsection{Introduction}

With the growing interest in the spin properties of
low-dimensional structures, particularly for spintronics and
quantum information applications, there is a need for
reliable experimental tools to obtain this information.  For
example, spin-orbit relaxation times can be determined by time-
resolved optical methods
\cite{awschalom01,awschalom01a,ensslin01} but an alternative
and complementary method is to use the weak antilocalization
(WAL) effect.  In metals, where it was thoroughly studied in
the eighties \cite{altshuler85,bergman84,chakravarty86}, WAL is
well understood, but for high mobility semiconductor structures
some refinement is needed if it is to become a reliable tool
for determining scattering parameters. 

For diffusion dominated transport the characteristic magnetic
field is $B_{tr} = \hbar/4eD \tau$ where {\it D} is the
diffusion constant and $\tau$ the scattering time. In metals
$B_{tr}$ is relatively large but in semiconductor samples it
can be very small: e.g. in the high mobility 2-dimensional
electron gas studied here, it is as small as 0.5 mT at zero
gate voltage. 
Weak localization (WL) effects
extend to fields several hundred times larger than this and
even the very narrow WAL peak extends well beyond B$_{tr}$. It
is not then valid to use low field approximations (which assume
$B \ll B_{tr}$)  to obtain experimental parameters
\cite{hikami80}. 
In this paper we will address the
issue of how to experimentally extract the phase-breaking 
($\tau_{\varphi}$) and spin-orbit ($\tau_{so}$)) time constants
under these conditions.  It will be experimentally established
that both the WL and WAL effects have the same orbital origin. 
Further, it will be shown that even when there is a crossover 
from weak to strong spin-orbit coupling, marked by a change 
from negative (WL) to positive (WAL) magnetoresistance 
as $\tau_{\varphi}/\tau_{so}$
increases,  the characteristic width of the peak continues to
be determined by $\tau_{\varphi}$. To determine $\tau_{so}$
accurately
requires that the whole curve, including the 
high field tail, be fitted.

\subsection{Experimental}

The sample studied was a high mobility, gated InGaAs
quantum well structure grown by chemical beam epitaxy on an InP
(100) substrate \cite{poole01}. 
This sample was of a particular interest because it exhibited
large spin-orbit effects.
 A cross-sectional layout view of the structure is shown in
Fig.1.  The quantum well is formed by 10 nm of
In$_{x}$Ga$_{1-x}$As ({\it x}=0.53) grown on an undoped 
InP buffer layer and separated from the Si-doped layer by a
30 nm spacer.  A rectangular Hall-bar sample, width 0.2 mm and
separation between adjacent potential probes 0.4 mm, was
fabricated using optical lithography and wet etching. A gold
gate was deposited on top of a 40 nm SiO$_2$ dielectric layer.

Experiments were performed in a He3 system (with temperatures
to
below 300 mK) in both perpendicular and tilted magnetic
fields. Measuring currents were 100 nA or smaller.  For precise

measurements in very small magnetic fields special attention
must be 
paid to the accuracy of the magnetic field.  A superconducting
magnet 
was used with the persistent switch was removed to ensure all
current 
delivered by the power supply passed through the magnet. The
magnet power 
supply (Oxford Instruments IPS120-10) had a stability and
reproducibility 
significantly better than 10$^{-5}$T.  To overcome the problem
of a trapped flux 
and the associated hysteresis in the magnet near zero field we
established 
a protocol for the magnetic field history which was calibrated
using a 
high sensitivity Hall probe. For  most  measurements the Hall
voltage from 
the sample was measured simultaneously and used to confirm the
accuracy of 
the magnetic field determined in this way.

Results of low-field Hall-effect measurements of the
concentration and mobility
as a function of gate voltage (V$_g$) are shown in Fig. 2.  The
concentration 
changes linearly with the gate voltage, as expected from a
simple capacitor 
model, indicating there was no electric-field dependent charge
accumulation 
between 2DEG and the gate.  The straight line in Fig. 2 is
calculated based on 
the parameters shown in Fig. 1 using an oxide thickness
$d_{ox}=40 nm$ and 
dielectric constants $\epsilon_{ox}$=3.9 and
$\epsilon_{InP}$=12.6 .  The 
electron mobility shown in Fig. 2 has a sub-linear gate voltage
dependence, changing from 8 to 1 m$^2$/Vs$^{-1}$ as the gate
voltage was reduced from 0 to -0.7 V.  This corresponds to a
characteristic
magnetic field B$_{tr}$ increasing  from 
0.5 mT at V$_g$ = 0 V  to 140 mT at V$_g$=-0.7 V.

Fig. 3 shows an example of the magnetoresistance (MR) measured
over a wide range of the magnetic field at several different
temperatures.  Four separate field regions can be
distinguished. 
At high fields (B $>$ 0.3 T) the Shubnikov-de Haas oscillations
are visible;
in an intermediate region there is a slow monotonic,
temperature dependent,
negative magnetoresistance. This parabolic term results from
the
electron-electron interaction effects
\cite{altshuler85,houghton82,minkov01} and will not be
discussed here.  Focussing on the low field region (B $<$
0.02T) 
both negative and positive MR components associated with
quantum interference corrections are seen.  It is
commonly accepted that the negative MR is due to the WL effect
and the central, very narrow,  dip to the WAL effect.
This dip, which appears only in samples where the spin-orbit 
scattering is strong, is so narrow that it could be used as an
absolute zero-field sensor, with a precision of better than
10$^{-5}$ T, in applications where it might be necessary to
compensate the Earth's magnetic field.

     The standard procedure to separate spin and orbital
effects is to make measurements with magnetic field tilted away
from the normal. Spin dependent terms, which depend on the
total magnetic field, then become enhanced relative to orbital
terms which depend only on the normal component of the field. 
Fig.4 shows MR traces for different tilt angles ($\theta$)
plotted as a function of the normal component $B\cos\theta$ 
\cite{footnote1} . If the WAL and WL components were to
originate 
from different mechanisms (e.g. WAL due to spin and WL due to 
orbital motion) a relative change in width of the two effects 
would be expected at different angles but in fact this is not 
so and the curves coincide. This implies that both the WL and
WAL 
effects depend only on the normal component of magnetic field
and 
that they both result from the orbital motion. It can be
concluded that any independent 
spin degree of freedom has been suppressed by the spin-orbit
coupling.  

\subsection{Weak anti-localization data in arbitrarily strong
magnetic fields}

     The magnetoresistance due to interference corrections
depends on
three characteristic field values  \cite{altshuler85,knap96}:

\begin{equation}
     B_{tr} = \frac{\hbar}{4eD\tau} , \; \;
     B_{so} = \frac{\hbar}{4eD\tau_{so}} \;\;\mbox{ and } \;\; 
    B_{\varphi} = \frac{\hbar}{4eD\tau_{\varphi}} 
\label{eq1x}
\end{equation}
where $D=l^2/2\tau$ is the diffusion coefficient,{\it l} is the
mean free path, and $\tau$, $\tau_{so}$ and
$\tau_{\varphi}$  are respectively the elastic scattering time,
the
spin-orbit relaxation time and the phase-breaking time.

     To extract these parameters from the MR traces it is
common to use the Hikami-Larkin-Nagaoka (HLN) equation
\cite{hikami80} but this is only valid for small magnetic
fields, $B \ll B_{tr}$ when the magnetic length
$l_B=\sqrt{\hbar/eB}$
is larger than the mean free path.  
In the high mobility sample considered
here $B_{tr}$ is very small (only $4.6\times 10^{-4}$T at
V$_g$=0 ) and $B_{so}$ and  $B_{\varphi}$ are even smaller
($0.9 \times10^{-4}$ and $7\times 10^{-6}$ T respectively).
As can be seen from Fig. 4 even the WAL peak extends beyond the
small field limit and it is therefore incorrect to use the HLN 
equation to extract these parameters.  The equation fails
because 
it was derived in the diffusion limit with sums over multiple
collisions replaced by integrals.  For fields larger than
$B_{tr}$,
when most closed path trajectories involve only a small number
of
collisions (as few as three), the sums have to be explicitly
evaluated. This situation was treated in ref.
\onlinecite{gasparyan85},
in the absence of spin-orbit effects, with the prediction that
there is a universal dependence ($\Delta\sigma(B)\sim
1/\sqrt{B}$) 
for the magnetoconductance at high fields. The more general
case,
when spin-orbit effects are included,  was considered by
Zduniak
{\it et al.} \cite{knap96}. Their expressions, which include 
both WL and WAL corrections to the conductivity, in arbitrary
magnetic
fields are:

\begin{equation}
     \Delta\sigma(B) = -K (e^2/\pi h) [ F(x,\beta_{s1}) +     
\frac{1}{2} F(x,\beta_{s2}) - \frac{1}{2} F(x,\beta_{\varphi})]
\label{eq2x}
\end{equation}

with

\begin{displaymath}
\begin{array}{rcl}
     F(x,\beta_i) & = &  x \sum_{n=0}^{\infty} 
     \frac{P_n^3}{1-P_n}  \\
 P_n(x,\beta_i) & = & (2/x)^{1/2} \int_{0}^{\infty} dt \,     
 exp(-(1+\beta_i)t(2/x)^{1/2} - t^2/2)L_n(t^2) \\ 
     L_n(t^2) & = & \sum_{m=0}^n (-1)^m \frac{n!}{(n-m)!}     
 \left[ \frac{t}{m!} \right] ^{2m}
\end{array}
\end{displaymath}
             
where L$_n$  are  Laguerre  polynomials, {\it i}=$\varphi$,{\it
s1} ,or {\it s2}, and
\begin{displaymath}
x = \frac{B}{B_{tr}} = \frac{4eBD\tau}{\hbar}, \;\;   
\beta_{\varphi} = \frac{\tau}{\tau_{\varphi}}, \;\;
\beta_{so} = \frac{\tau}{\tau_{so}}, \;\;
\beta_{s1} = \beta_{\varphi} + \beta_{so},\;\;
\beta_{s2} = \beta_{\varphi} + 2\beta_{so},
\end{displaymath}

Here (as is discussed below) an extra, empirical, coefficient
K has been 
introduced as compared to ref. \onlinecite{knap96} to allow
good fitting to the experimental data 
over the whole range of magnetic fields.  
To reduce computation time when fitting data the function
$F(x,\beta _i)$
was calculated using 2000 Laguerre polynomials and stored
numerically as a  matrix of $F_{ij}=F(x_i,\beta_j)$ on a
semi-logarithmic mesh. 
Values between defined points $(x_i,\beta_j)$ and
$(x_{i+1},\beta_{j+1})$ were
determined by linear interpolation. 

Although the calculated quantity is $\Delta \sigma$ that
measured is $\rho_{xx}$. 
Even in the absence of any interference corrections  
$\sigma_{xx}$ =$\rho_{xx}/(\rho_{xx}^2+\rho_{xy}^2)$ has a
small quadratic field 
dependence, which, in high mobility samples, cannot be ignored.
It can be corrected 
for by comparing the calculated quantity  $\Delta
\sigma_{WL}(B)$ not 
with $\Delta \sigma_{xx}$ but rather with  
$\Delta (1/\rho_{xx})= 1/\rho_{xx}-1/\rho_0$, which 
classically has no field dependence.

Figure 5 gives an example of experimental data of $
\Delta(1/\rho_{xx})$
which compares with calculated values of $\Delta\sigma$ 
obtained from 
Eqn.(\ref{eq2x}) with K = 1.  A reasonably good fit to the low
field part of the 
experimental data can be obtained with $\beta_{\varphi}$ =
0.005 and 
$\beta_{so}$ = 0.38 but the calculated curve deviates
significantly from 
the data at higher fields. In high-field region (B/B$_{tr}>1$),
where 
universal behaviour is expected \cite{gasparyan85,knap96}.  The
high field tail can be 
fitted with a range of values of $\beta_{\varphi}$ and 
$\beta_{so}$ 
provided only that they are small ($\beta_{\varphi}$, 
$\beta_{so} < 0.01$). 
Any adequate fit to the high field tail, however, leaves a
large discrepancy 
in the low-field region (B/B$_{tr}<1$). Conversely, although
the shape of 
the WAL peak depends mainly on $\beta_{\varphi}$ the turn-over
from WAL to
WL behaviour is determined essentially by $\beta_{so}$. Values
of $\beta_{\varphi}$ 
 and $\beta_{so}$ large enough to describe low field dependence
properly are 
then too large to fit the high field part of the data. 
Exactly the same problem is also evident in other works, e.g.
in Ref. \onlinecite{knap96} 
where universal behaviour of the magnetoresistance at high
fields is reported and fitted using 
reasonable  parameters but only at the expense of poor fits at
low fields.  

The problem of fitting the magnetoresistance associated with
WAL effects in semiconductor structures, over a wide range of
magnetic fields, is well known. Weak localization in
semiconductors is more complex than in metals because of high
electron mobilities and because new mechanisms involving spin
orbit effects appear.  One purpose of this paper is to alert
theorists to this issue. As noted above, papers that consider
WL effects in arbitrary magnetic fields, e.g. Ref.
\onlinecite{gasparyan85, knap96}, are unable to adequately
describe the experiments. Despite this it is possible to obtain
estimates of the phase-breaking and spin-orbit scattering times
from   experimental MR curves that may have systematic errors
but will nevertheless correctly reproduce gate voltage and
temperature dependences. One commonly used procedure is to fit
only the  low field  part of the MR using the HLN expression
\cite{hikami80}. In this paper we use the more elaborate
expression Eq.\ref{eq2x} (with K=1)  which coincides with HLN
formula at small fields. Secondly, we fit data over the whole
range of the magnetic field by introducing the extra, empirical
coefficient K. Because we can provide no theoretical
justification for the coefficient K, we present results for
$\tau_{\varphi}$ and $\tau_{so}$ determined with both K=1 and
K allowed to vary.

The fits to the low field data with K=1 (see Fig.5) not only
failed to describe the high field tail but also gave
unreasonably large values for the spin-orbit parameter
$\beta_{so}$. For example the value of 0.38 used in Figure 5
corresponds to the unphysical value of approximately one for
the parameter $\beta_{s2}$ in Eqn.(\ref{eq2x}).  Examples of
fits with K allowed to vary are shown in Fig.6.  In this case
fits for all temperatures gave K=2.1$\pm$0.1. For more negative
gate voltages 
the high field data had essentially the same, universal,
behaviour and could again be 
adequately fitted with K=2 although with an increased
experimental uncertainty. To make the  
comparisons of $\beta_{\varphi}$ and $\beta_{so}$ more
meaningful it was therefore decided to fix K at 2 with a
corresponding reduction in the uncertainty with which the other
parameters could be determined.

  With the empirically introduced 
coefficient K it was possible to achieve satisfactory fits to
the data, over 
the whole field range, for all temperatures and gate voltages.
We note that K does 
not appear to be a universal coefficient; in other samples
\cite{unpublished1} 
values of K smaller than 2  were needed to fit the data.  The
failure of 
the theory with K = 1 raises questions about other fitting
procedures commonly
used in the literature, in particular the HLN formula which, at
low fields, is 
equivalent to Eqn.2 with K=1  \cite{knap96}.   Fitting to just
the low field
(WAL) region with K=1, i.e. relaxing the requirement that the
high field behaviour 
be adequately described (see Fig. 5), gives values for the
parameter 
$\beta_{\varphi}$  several times smaller and $\beta_{so}$
several times larger than 
those obtained with K=2. Fitting to the low field region using
the HLN equation 
gave very similar parameters but with even larger deviations at
high
fields. 

While it is common to offset the theoretical curves to have a
value of
$\Delta \sigma$ =0 at B=0 (as shown for example in Fig. 5) the
theoretical 
values given by Eqn. \ref{eq2x} tend to zero in the limit of
high 
magnetic field where both the WL and WAL effects are fully
quenched. This means that fits made without any offset (for
example those shown
in Fig.6) determine the absolute values of the interference
correction to the 
conductivity. The temperature dependence seen in Fig. 6 shows
a universal behaviour at high fields incresing with the same
slope 
 but  low field (WAL) 
behaviour has a strong temperature dependence. As a 
function of temperature $\tau_{\varphi}$ is expected to change
but 
$\tau_{so}$ remain constant \cite{altshuler85,bergman84}. It is
often assumed, 
when WAL is present, that the low field dependence is
determined by 
$\tau_{so}$ and the high field with $\tau_{\varphi}$ . This
would 
imply a temperature dependent high field region but unchanged
WAL peak, 
in direct contrast to what is observed experimentally. The 
calculated fits (solid lines in Fig. 6) did confirm this point.

The changing amplitude of the WAL peak corresponds to a
temperature dependent phase-breaking time $\tau_{\varphi}$  and
the ``universal'' high field slope corresponds to a value of
$\tau_{so}$ that is essentially independent of the temperature.
This 
happens when there is strong spin-orbit scattering, that is  
$\tau_{so} < \tau_{\varphi}$.

     We conclude therefore, perhaps counterintuitively, that
the orbital motion (the phase-breaking time) determines
the width of the central WAL peak, but the strength of the
spin-orbit scattering ($\tau_{so}$) controls
the high field ``universal'' behaviour. This behaviour is
reflected
in the HLN formalism \cite{hikami80} which although not
strictly
valid for the high mobility sample measured here reflects
the correct physics and has the advantage it can be treated
analytically. For small B ($\ll B_{\varphi}$) it gives 

\begin{equation}
     \Delta\sigma =  \frac{e^2}{\pi h} \frac{1}{24}
 \left( \frac{B}{B_\varphi} \right) ^2 
\Theta(\frac{\tau_\varphi}{\tau_{so}}),
\label{eq3}
\end{equation} 
where
\begin{displaymath}
     \Theta(\frac{\tau_\varphi}{\tau_{so}}) =
\frac{1}{(1+\tau_\varphi /\tau_{so})^2} + 
\frac{1}{(1+2\tau_\varphi /\tau_{so})^2} -1
\end{displaymath} 
The dimensionless function $\Theta$
depends only on the ratio $\tau_{\varphi}/\tau_{so}$. For
$\tau_{so} \rightarrow \infty$, corresponding to pure WL,
$\Theta$=1 and the standard expression, given for example in
ref. \onlinecite{altshuler85}, is recovered. In the opposite
limit however, $\tau_{\varphi}/\tau_{so} \gg 1$, the absolute
value 
of $\Theta$ is still equal to one but the sign changes. The
characteristic width of the peak, in both limits, is therefore
determined
by $\tau_{\varphi}$, the amplitude by the the ratio
$\tau_{\varphi}/\tau_{so}$.
 The change of  sign for a ratio $\approx$ 0.3 not one,
reflects the
fact that the spin-orbit interaction is three dimensional in 
nature with three spin components to relax compared with one
for the scalar 
phase breaking process. 

     The WAL peak is therefore so narrow because the width is
determined not by $\beta_{so}$ but rather by $\beta_{\varphi}$
which can be
extremely small in high mobility structures ( e.g. $7\times
10^{-6}$
T here at V$_g$=0). In the absence of spin-orbit scattering,
there would be a WL peak, with the same extremely narrow width
but of the opposite sign.

\subsection{Discussion}

In this section the temperature and gate voltage
(concentration) dependence of the phase-breaking time and spin-
orbit interaction constant will be discussed.  In the absence
of a theory
that can satisfactorily describe the magnetoconductivity over
the whole
field range we present values of $\tau_{\varphi}$ and
$\tau_{so}$ detrmined
using both K=2  and K = 1 as it discussed earlier.  Figure 8
shows the phase breaking time $\tau_{\varphi}$ as 
a function of temperature extracted by fitting the data such as
that shown 
in Fig. 7. For both K=1 and K=2 the behaviour is similar with
a linear dependence 
at higher temperatures and an essentially  reduced slope below
1 K.   
The solid line shows a theoretical limit due to the
electron-electron scattering 
based on a Fermi-liquid model \cite{altshuler85,altshuler82}:

\begin{equation}
     \frac{1}{\tau_{\varphi}}=\frac{k_B T}{\hbar}\frac{\pi
G_0}{\sigma_0}ln\left(\frac{\sigma_0}{2\pi G_0}\right)     
\label{eq4}
\end{equation}
with $G_0=e^2/(\pi h)$, and where $k_BT\tau/\hbar \ll 1$. It
should
be noted that in the literature an empirical coefficient of
order 2 is
often introduced to bring the experimental data into better
agreement with Eq. \ref{eq4}  \cite{polyanskaya97, minkov01}. 
 This
model works well in metals, where Fermi-energy is large and the
electron gas can be considered as being very uniform
\cite{altshuler85,bergman84}, but a saturation of
$\tau_\varphi$ is usually 
reported  at low temperatures (see e.g. ref.
\onlinecite{bergman84}).
Similar behavior is observed in Fig. 7: at high temperatures
 $\tau_{\varphi}$ detrmined using K=2 increases linearly with
decreasing temperature
with the expected slope and there is a saturation below 1K. For
K = 1 the behaviour
is qualitatively similar although less pronounced. In both
cases the values at 
high temperatures is a factor of two or three smaller than
expected in Fermi-liquid model. The saturation 
below 1K suggests some additional phase-breaking mechanisms
limit ${\tau_\varphi}$. 
Possibilities include inhomogeneous distribution of alloy
composition, interface
roughness or doping concentration variations
\cite{minkov01a,germanenko01, mathur01}. 
In high mobility samples such as that studied here small
fluctuating magnetic fields
may also be playing a role. The maximum value of
$\tau_{\varphi}\simeq 100 ps$  corresponds to
$B_\varphi\simeq$0.012 mT.  
This is anextremely small field, several times smaller than the
Earth's magnetic field, so 
any fluctuating or micro-scale effective magnetic field of this
magnitude would 
affect the very narrow WAL peak and appear as a phase-breaking
mechanism.
Permanent {\it dc} magnetic fields, such as the Earth's field,
would lead only
to an arbitrary shift in the position of the peak and in-plane
components of the 
field would also have no effect. ({\it cf} Fig. 4.)

    While any detailed analysis of the mechanisms of phase
breaking is 
beyond the framework of this paper but it can be concluded that
the 
WAL effect provides a useful tool for determining and
controlling the 
phase breaking time. In the sample used here the value of about
100 $ps$
corresponds to a phase breaking length $l_\varphi = 20-40
\mu$m.   

Gate voltage dependence of the magnetoconductivity is shown
in Fig. 8. In this figure all the curves shifted vertically to
coincide
at B=0.  Rather surprisingly, when plotted in this way,
universal behaviour is observed at low magnetic fields ($B
<B_{tr}$)
 with the WAL peak for different gate voltage data
collapsing onto a single logarithmic curve.   Indeed, the low
field WAL peak in Fig.8 now shows a similar kind of $\sim
\ln(B)$ dependence seen in the high-field (WL) part but with
the opposite sign. 

The results from fitting this data are plotted in Fig. 9 as a
function of the 
conductivity to be able to compare the results with 
Fermi-liquid model (Eq. 4). Again two values of K have been
used and in both cases the variation
of $\tau_{\varphi}$ is much slower than is predicted
theoretically by the 
Fermi-liquid model Eq. (4). While it is not clear which of the
curves is correct
they both lie below the theoretical one and have a slower
dependence on conductivity.  
This may be associated with the fact that the measurements were
made at the lowest
temperature and therefore just be reflecting the saturation
observed in the temperature
dependence (Fig. 7).

As noted above the width of the WAL feature depends on
$\tau_\varphi$ but the amplitude and the transition to the high
field tail also depends on $\tau_{so}$. The physics describing
the damping 
of the spin-orbit interaction is more complicated than for the
dephasing. 
To describe the WAL effect a spin-dependent vector potential is
required 
with a three dimensional character
\cite{lyandageller98,pikus95, iordanskii94}. 
Different spin-orbit relaxation mechanisms are not additive and
more
complicated expressions, with more fitting parameters, should
be
used to describe experiments. If, however, only one spin-orbit
mechanism dominates, as seems to be the case here, a single
scalar parameter $\tau_{so}$ should suffice which can then be
treated
on the same footing as $\tau_\varphi$.  The values of
$\tau_{so}$ 
determined from fits to the field dependences as a function of
density (Fig. 8),
are plotted in Fig. 10 (a), again for K=1 and K=2.  The
spin-orbit relaxation 
time is significantly smaller than $\tau_\varphi$  (and only a 
few times larger than transport relaxation time). For K=2 
$\tau_{so}$
 increases from 12 to 19 ps as the concentration decreases from
3.5 to
$1.5\times10^{11}$cm$^{-2}$; for K=1 case the deduced values of
$\tau_{so}$ are 
even smaller. Small values of $\tau_{so}$ are consistent with
the strong
spin-orbit coupling in the InGaAs which means that any elastic
scattering event has a high probability of also involving spin
scattering.

     Two major spin-orbit scattering mechanisms are expected
for 2DEG systems such as that considered here: the Dresselhaus
term, associated with the bulk zinc-blend crystal inversion
asymmetry and the Rashba term, associated with a built-in
electric field. \cite{pikus96}  To distinguish which mechanism
dominates 
it is helpful to consider the dependence of
$B_{so}=\hbar/(4eD\tau_{so})$ 
on electron concentration \cite{pikus96, dresselhaus92,
chen93}. 
In particular the Dresselhaus term is expected to increase with

increasing carrier density. For example, in a GaAs/AlGaAs
heterostructure 
 a  quadratic increase of $B_{so}$ with density is predicted
and 
was observed experimentally \cite{dresselhaus92} . Fig. 10 (b)
shows $B_{so}$ 
as a function of electron concentration. (Note that though
$B_{so}$ is inversely 
proportional $\tau_{so}$ the stronger density dependence of D
means $B_{so}$ 
also decreases with density). The approximately inverse
parabolic dependence
that is observed cannot be attributed to the Dresselhaus
mechanism. 

The Rasba term, which appears in asymmetric quantum wells,
contributes
a term $H_R=\alpha {\bf [\bar\sigma \times \bar k]}_z$ to the
Hamiltonian
with the coefficient $\alpha$ proportional the expectation
value of the
electric field in the well. In the literature the
role of interfaces in the Rashba mechanism is somewhat
controversial. Within the effective mass approximation the
expectation value 
of a (smooth) potential gradient integrated over the whole
space is always 
zero \cite{pikus95,gerchikov92}. More generally, the interfaces
should 
be treated separately and with contributions that may be as
large (or even larger)
as that from the quantum well \cite{gerchikov92}. The two
interfaces in a 
quantum well usually have different properties, because of
differences in
the growth process. Changing the gate voltage will therefore
not only change 
the average built-in electric field in the well but also the
relative interaction 
of the electrons with the different interfaces. 

     The density dependence seen in Fig. 10(b) is of the
opposite sign to 
that expected for a simple triangular confining potential.
Simulations have
shown, however, that this kind of functional dependence might
be explained
qualitatively by the built-in electric field  \cite{chen93}
(excluding the effect of the interfaces) provided the
background doping of the buffer layer (which contributes
2.2x10$^{11}$ cm$^{-2}$ carriers to the quantum well) is also
taken
into account. However, the magnitude of the effect is larger
than expected and a more detailed study, outside the scope of
the present paper, is needed to settle this point.

\subsection{Conclusions}

Interference corrections to the conductivity have been studied
in a high-mobility {InGaAs/InP} quantum well with the
particular
intent of examining the WAL effect and refining the procedures
needed to establish it as a tool for gaining information about
phase-breaking and spin-orbit coupling processes.  When the
magnetoresistance 
was examined over a wide range of magnetic fields $0\leq
B/B_{tr}\leq 100$ 
it was found that functional dependence given in ref.
\onlinecite{knap96}
could not adequately describe the data. Reasonable fits could
be obtained
by introducing an empirical amplitude factor $\simeq 2$. 
The reason for this disagreement is not understood and
it would obviously be interesting to make similar
measurements and analysis, over a wide field range, in other
semiconductor
systems.  One possible reason is that the spin-orbit coupling
is
sufficiently strong in this particular InGaAs QW sample that
the 
theory \cite{knap96} is starting to fail because the condition
 $\tau_{so}\gg \tau$ is not well satisfied.  In
this case an alternative approach, based perhaps on a spin-
dependent vector potential  \cite{lyandageller98} needs to be
developed.

Despite this disagreement several conclusions can be drawn from
this study, 
summarized as follows. The WAL and WL effects both have an
orbital origin and
depend only on the perpendicular component of the magnetic
field. For 
 $\tau_{\varphi}/\tau_{so}/ \ll 1 $ the central, low field
peak, has WL 
character and for $\tau_\varphi/\tau_{so}/\gg 1$ WAL character,
but in both 
cases the width of the low-field  peak is determined only by
$\tau_\varphi$. The high field
dependence is universal with the cross-over from the low field
behaviour determined
by the ratio $\tau_{\varphi}/\tau_{so}$.
 
 The spin-orbit scattering time is small, between about 12 and
18 ps, and 
only weakly dependent on the electron concentration.  As has
been found in many other 
studies the experimentally determined dependence of
$\tau_\varphi$ on 
temperature and gate voltage cannot be satisfactorily 
described 
by Fermi-liquid theory, some additional phase-breaking
mechanisms appear to be present.

Overall, we have demonstrated it is possible to use gate
voltage to control the 
strength of the spin-orbit interaction.  It was also shown  
that the magnetoresistance associated with the quantum
interference corrections provides a powerful tool for
controlling and studying
the interplay between the phase-breaking time and spin-orbit
coupling in low-
dimensional structures.  However, a theoretical understanding
of these
effects is still not complete, particularly for arbitrary
magnetic field strengths
 and strong spin-orbit coupling.

\subsection{Acknowledgements}

We would like to acknowledge fruitful discussions with: Yu.
Lyanda-Geller, Geof Aers, Boris Narozhny, Chandre Dharma-
wardana and Sergei Dickmann.

\newpage


\begin{figure}[tbp]
\includegraphics[height=70mm,clip=false]{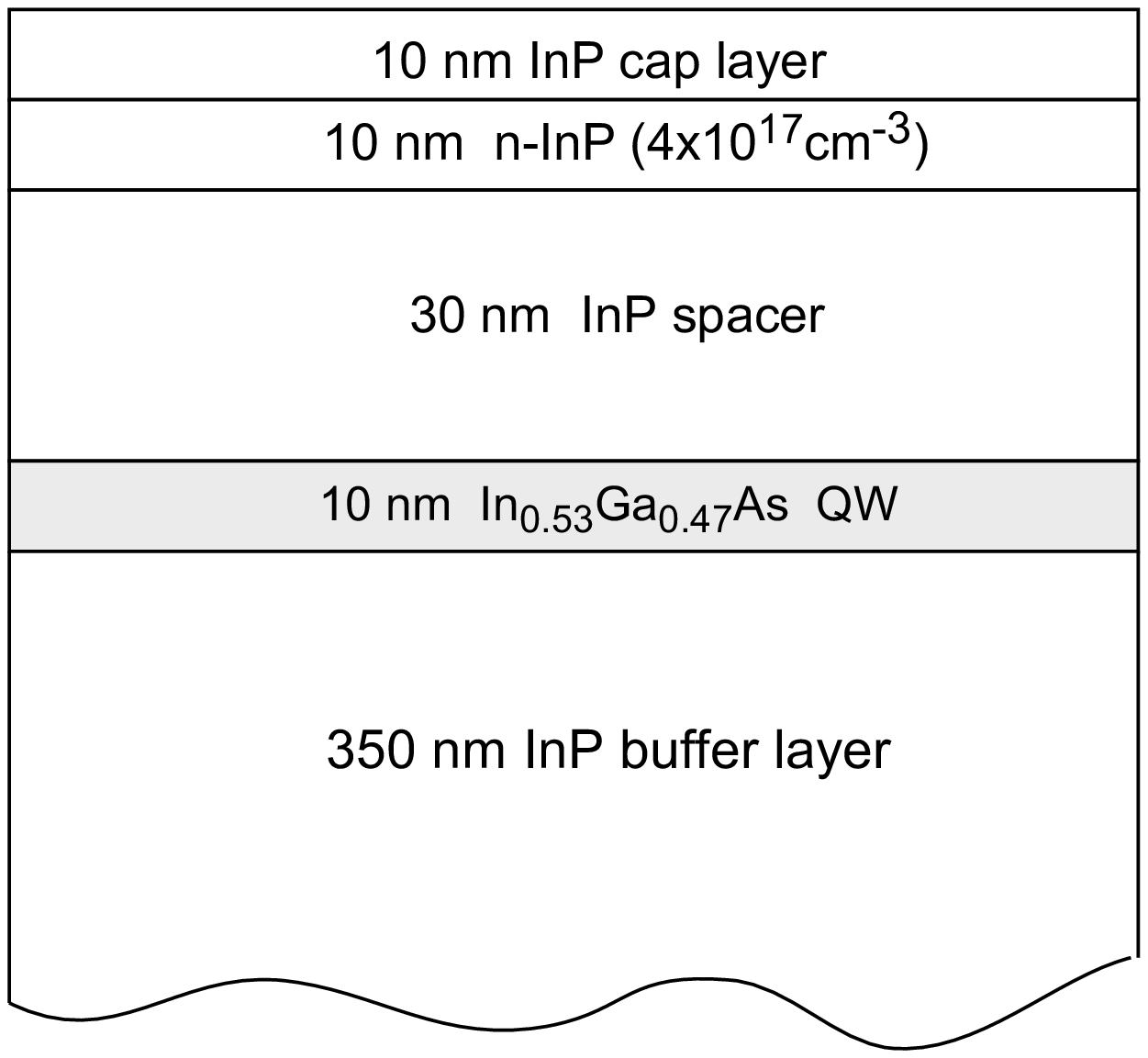}
\caption{Cross-sectional layout view of the InGaAs/InP quantum
well structure.} 
\label{fig1}
\end{figure}

\begin{figure}[tbp]
\includegraphics[height=70mm,clip=false]{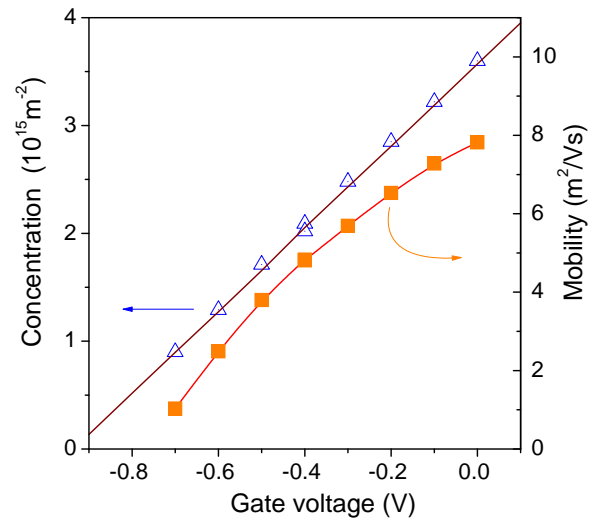}
\caption{Results of Hall-effect measurements of the electron
concentration and mobility {\it vs} gate voltage.  } 
\label{fig2}
\vskip0.5cm  
\end{figure}

\begin{figure}
\includegraphics[height=70mm,clip=false]{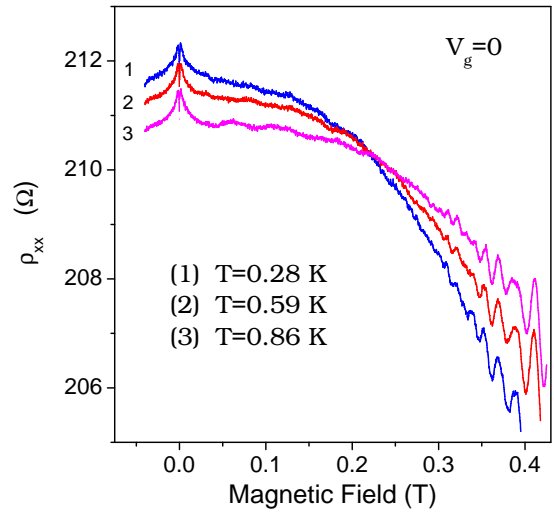}
\caption{Magnetoresistance traces from the InGaAs/InP quantum
well structure at different temperatures for a wide range of
the magnetic fields.} 
\label{fig3}
\vskip0.5cm  
\end{figure}

\begin{figure}
\includegraphics[height=70mm,clip=false]{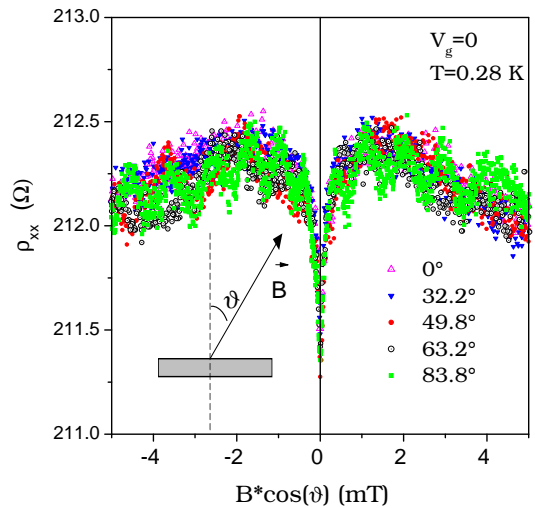}
\caption{Low field magnetoresistance attributed to quantum
interference corrections in tilted magnetic fields plotted as
a function of the normal component of magnetic field.}
\label{fig4}
\vskip0.5cm  
\end{figure}

\begin{figure}
\includegraphics[height=70mm,clip=false]{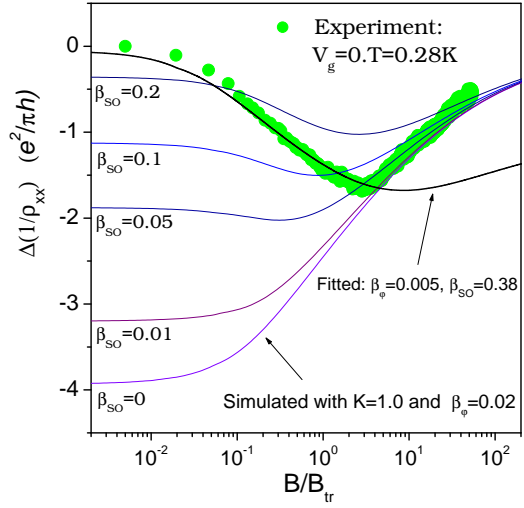}
\caption{Conductivity plotted against normalized magnetic
field.  Points are experimental data at V$_g$=0 and T=0.28 K.
Lines are simulated dependencies from eq.(2), all with K=1. } 
\label{fig5}
\vskip0.5cm  
\end{figure}

\begin{figure}
\includegraphics[height=70mm,clip=false]{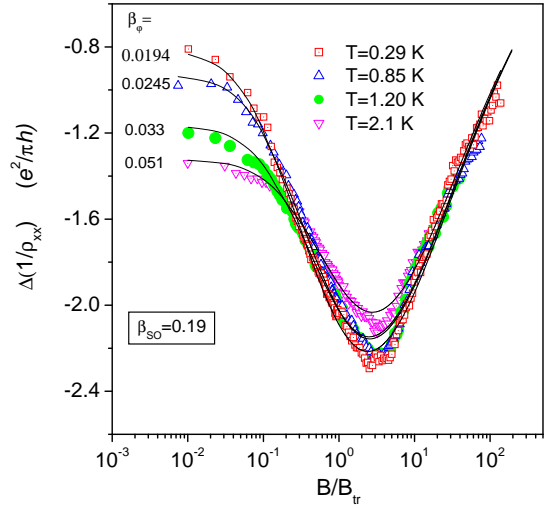}
\caption{Magnetoconductivity, at different temperatures and
with V$_g$=0, plotted against normalized magnetic field.  Lines
are fitted dependences with Eq.(2) using K=2. The experimental
data are offset to coincide with the theoretical curves which
approach zero in strong magnetic fields.  A universal behavior
is observed in high magnetic field region. Amplitude of the WAL
peak at B=0 strongly depends on temperature.} 
\label{fig6}
\vskip0.5cm  
\end{figure}

\newpage

\begin{figure}
\includegraphics[height=70mm,clip=false]{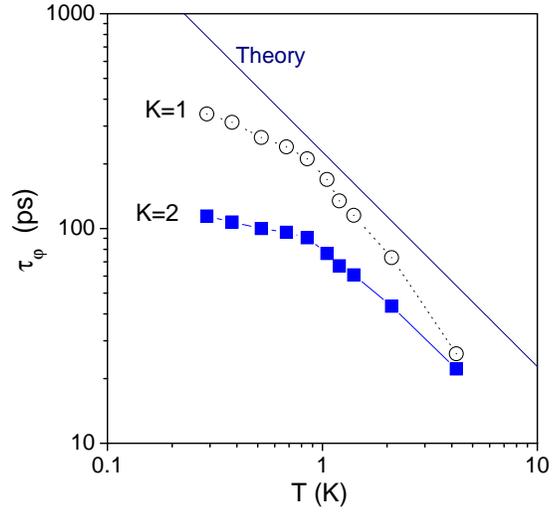}
\caption{Phase-breaking time $\tau_{\varphi}$, as a function of
temperature, extracted by fitting data in Fig.\ref{fig6} with
K=2 (solid squares) and K=1 (open circles). Solid
line is a theoretical limit due to the electron-electron
scattering (eq. 4).      } 
\label{fig7}
\vskip0.5cm  
\end{figure}

\newpage
\begin{figure}
\includegraphics[height=70mm,clip=false]{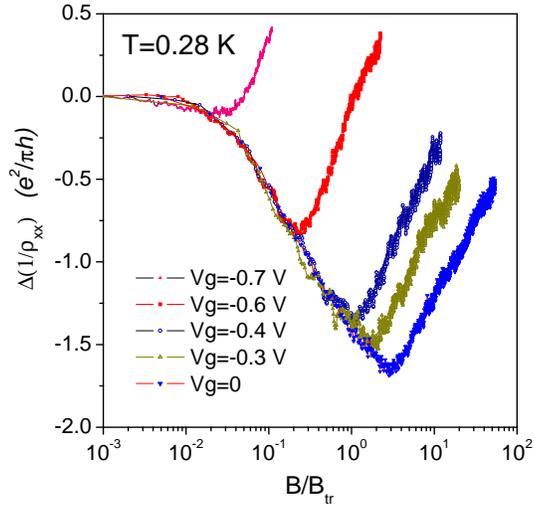}
\caption{Magnetoconductivity plotted against normalized
magnetic field for different gate voltages. The experimental
curves are offset to have the same value at B=0. } 
\label{fig8}
\vskip0.5cm  
\end{figure}

\begin{figure}
\includegraphics[height=70mm,clip=false]{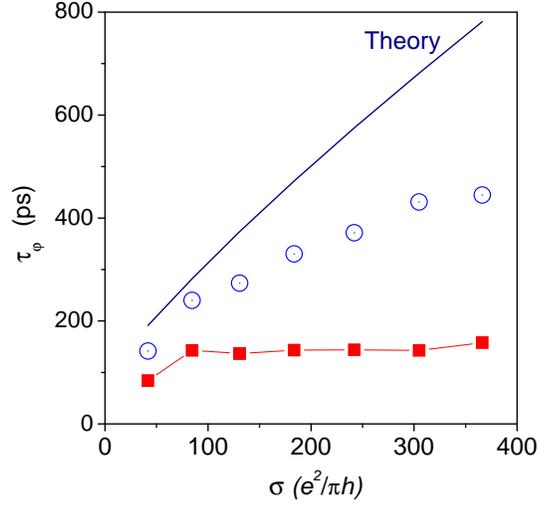}
\caption{Phase braking time vs conductivity. Line is
theoretical prediction based on Fermi liquid model eq.
\ref{eq4}, points are experimental results obtained by fitting 
data in Fig. 8 with eq. \ref{eq2x} using K=1 (open circles) 
and K=2 (solid squares).}
\label{fig9}
\vskip0.5cm  
\end{figure}

\begin{figure}[tbp]
\includegraphics[height=70mm,clip=false]{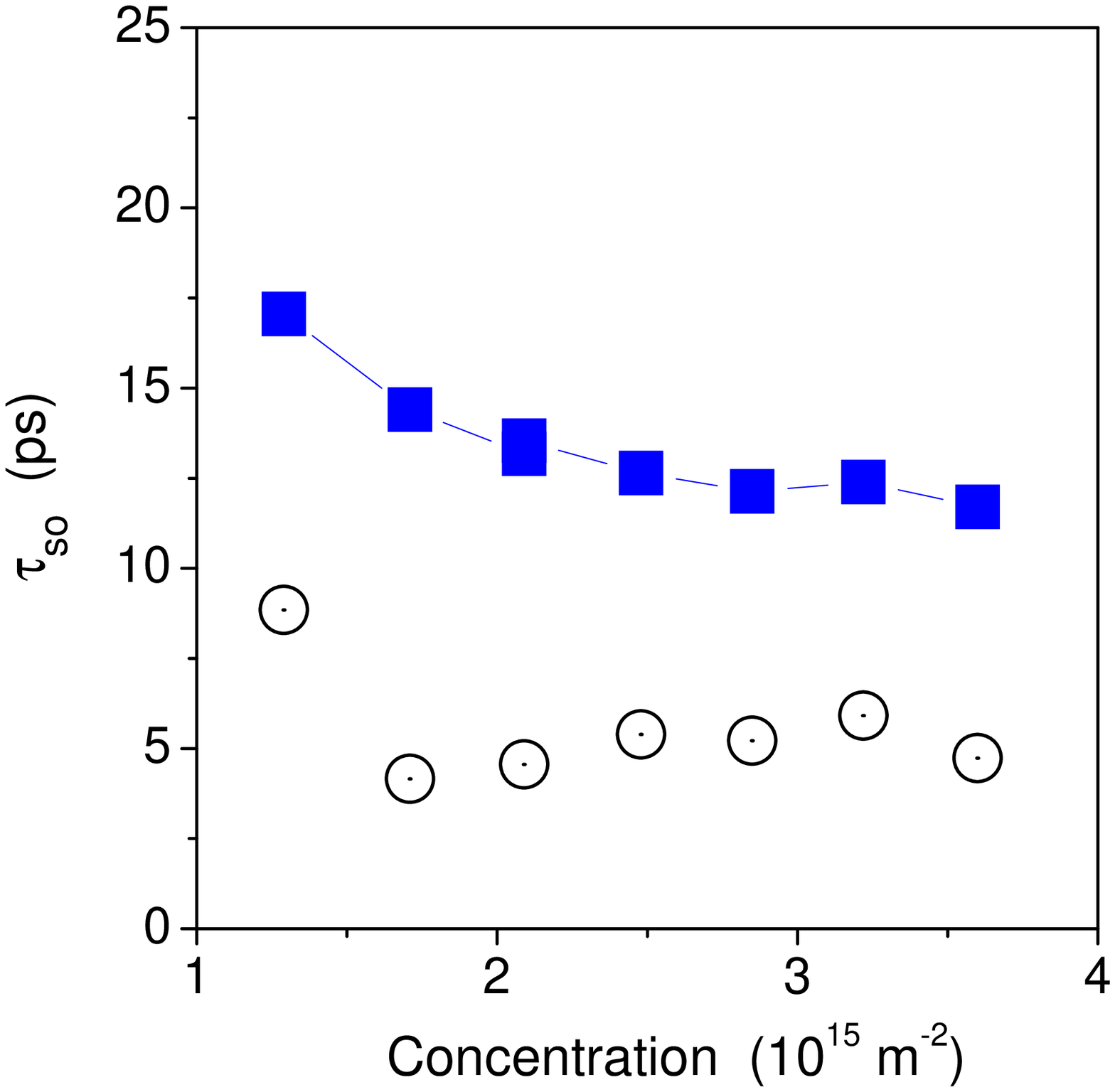}
\includegraphics[height=70mm,clip=false]{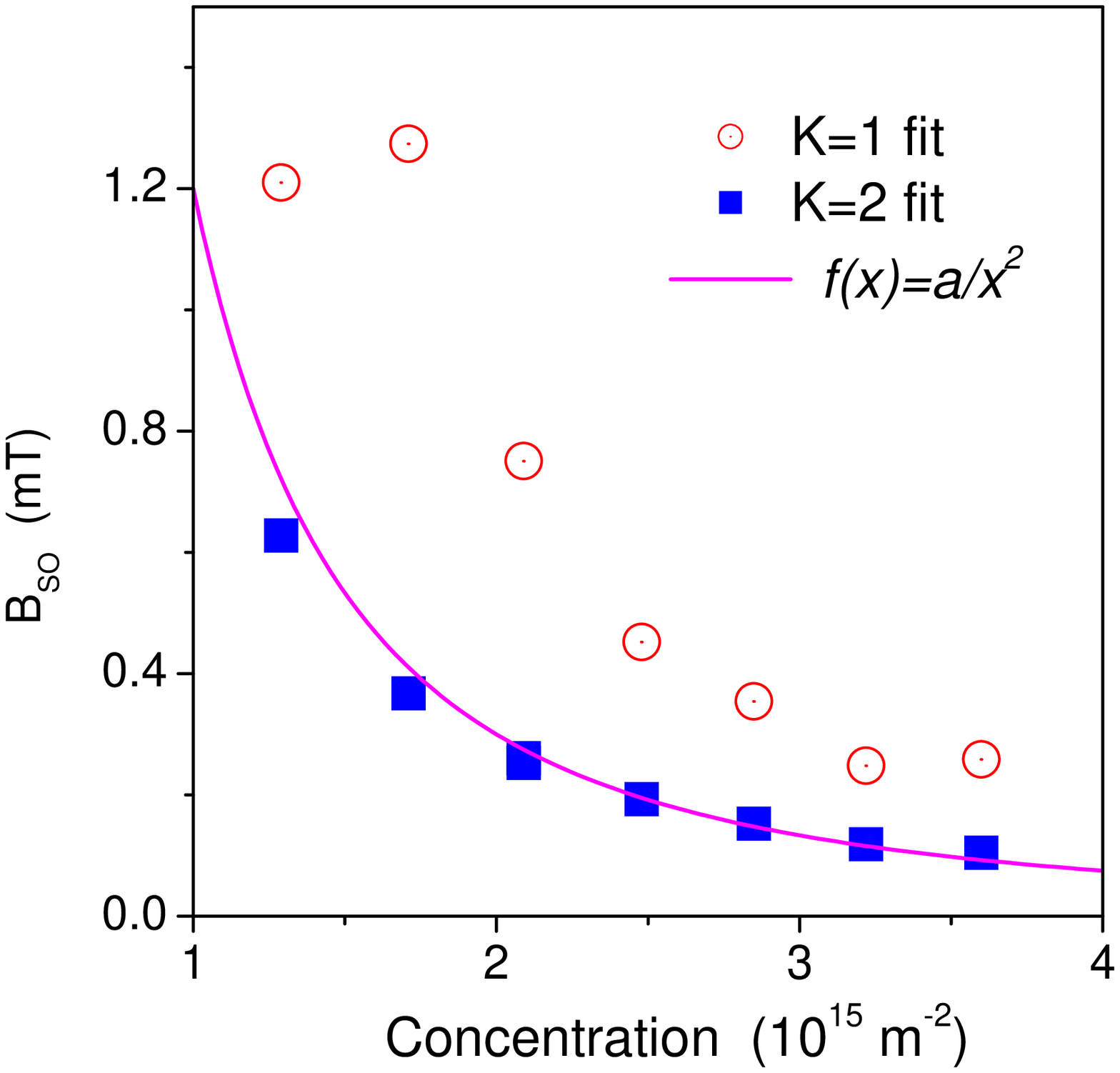}
\caption{ (a) Spin-orbit scattering time as a function of the
2DEG
concentration determined from fits to the data in 
Fig.\ \ref{fig8} using K=1 (open circles) and K=2 (solid
squares). \\
(b) Characteristic magnetic field value B$_{SO}$ as a
function of the electron concentration calculated on the basis
of the
data in Fig. 10 (a) and Fig.\ \ref{fig2}.  Solid line is
a fit proportional to inverse square of the electron
concentration. 
 } 
\label{fig10}
\end{figure}


\begin{references}
\bibitem{footnote1} It should be noted this data was taken at
the lowest temperatures
and using a measuring current of only 30 nA to avoid possible
heating problems. The noise 
is therefore larger than for other data.
\end{references}
\end{document}